\documentclass[aps,prb,twocolumn,amsmath,longtitle,superscriptaddress,floatfix]{revtex4-1}

\usepackage{amsmath}
\usepackage{amsfonts}
\usepackage{bm}
\usepackage{epsfig}
\usepackage{graphicx}
\usepackage{graphics}
\usepackage{url}
\usepackage[english]{babel}
\usepackage{dcolumn}
\usepackage{color}

\begin{document}

\title{Quantum dynamics of optical phonons generated by optical excitation\\ of a quantum dot: A Wigner function analysis}

\author{Daniel~Wigger}
\affiliation{Institut f\"ur Festk\"orpertheorie, Universit\"at M\"unster, Wilhelm-Klemm-Str.~10, 48149 M\"unster,
Germany}

\author{Helge~Gehring}
\affiliation{Institut f\"ur Festk\"orpertheorie, Universit\"at M\"unster, Wilhelm-Klemm-Str.~10, 48149 M\"unster,
Germany}

\author{V.~Martin~Axt}
\affiliation{Theoretische Physik III, Universit\"at Bayreuth, 95440
Bayreuth,
Germany}

\author{Doris~E.~Reiter}
\affiliation{Institut f\"ur Festk\"orpertheorie, Universit\"at M\"unster, Wilhelm-Klemm-Str.~10, 48149 M\"unster,
Germany}

\author{Tilmann~Kuhn}
\affiliation{Institut f\"ur Festk\"orpertheorie, Universit\"at M\"unster, Wilhelm-Klemm-Str.~10, 48149 M\"unster,
Germany}

\date{\today}
\begin{abstract}
The study of the fundamental properties of phonons is crucial to understand
their role in applications in quantum information science, where the active
use of phonons is currently highly debated. A genuine quantum phenomenon
associated with the fluctuation properties of phonons is squeezing, which is
achieved when the fluctuations of a certain variable drop below their
respective vacuum value. We consider a semiconductor quantum dot in which the
exciton is coupled to phonons. We review the fluctuation properties of the
phonons, which are generated by optical manipulation of the quantum dot, in
the limiting case of ultra short pulses. Then we discuss the phonon
properties for an excitation with finite pulses. Within a generating function
formalism we calculate the corresponding fluctuation properties of the
phonons and show that phonon squeezing can be achieved by the optical
manipulation of the quantum dot exciton for certain conditions even for a
single pulse excitation where neither for short nor for long pulses squeezing
occurs. To explain the occurrence of squeezing we employ a Wigner function
picture providing a detailed understanding of the induced quantum dynamics.
\end{abstract}


\maketitle

\section{Introduction} \label{sec:intro}
Phonons and their interaction with the electronic degrees of freedom are
omnipresent in solid state devices. Typically associated with heat, noise or
dissipation, nowadays phonons are becoming actively used, which is the
foundation of the emerging field of phononics \cite{volz2016nan}. Examples
are the use of phonons in the form of strain pulses to manipulate the lasing
properties of semiconductor structures
\cite{bruggemann2011las,czerniuk2014las} or the application of phonons in the
form of surface acoustic waves to control the dynamics in quantum dots
\cite{stotz2005coh,volk2010enh,fuhrmann2011dyn,weiss2014dyn,gustafsson2014pro}.
The generation of  coherent phonons in semiconductor nanostructures has been
studied \cite{kerfoot2014opt,nakamura2015inf} and also phonon lasers have
been proposed \cite{kabuss2012opt}. To explore genuine quantum features of
phonons it is interesting to study their fluctuation properties and in
particular the emergence of squeezing. Squeezing refers to the reduction of
fluctuations of a certain variable below their vacuum level. However, one has
to keep in mind that the Heisenberg uncertainty principle has to be
fulfilled, which results in increased fluctuations of the conjugate variable.
For photons, squeezing is well explored
\cite{dodonov2002non,polzik2008qua,drummond2013qua} and is already used in
applications, e.g., to detect gravitational waves at the LIGO experiment
\cite{goda2008aqu}. The prospect of finding squeezing also in a mechanical
system like phonons, and in particular in semiconductor systems, has
triggered a lot of theoretical
\cite{janszky1990squ,hu1996squ,sauer2010lat,papenkort2012opt,zijlstra2013squ}
and experimental work
\cite{garrett1997vac,misochko2000imp,johnson2009dir,esposito2015pho}. In this
paper, we will discuss the emergence of squeezing for phonons generated by
optically exciting a semiconductor quantum dot.

To be specific, we will study the fluctuation properties of optical phonons
which result from an excitation with finite pulses. We will show that the
pulse duration is indeed a crucial parameter for phonon squeezing. The
theoretical background is introduced in the next Sec.~\ref{sec:theory} and
the results for various types of excitation conditions are presented in
Sec.~\ref{sec:results}. The paper ends with some concluding remarks in
Sec.~\ref{sec:conclus}.

\section{Theory} \label{sec:theory}
When searching for non-classical phonon states, it is convenient to use a
system, where the electronic part is as simple as possible such that one can
focus on the phonon properties. One such system is a self-assembled
semiconductor quantum dot (QD) which, under certain conditions, constitutes
an electronic two-level system. When the electronic configuration in the QD
system changes, in particular by optical excitation with laser light, the
lattice reacts to this change by creating phonons. We will show that this can
also affect the phonon fluctuations which opens up the possibility to
manipulate these fluctuations by changing the excitation conditions.

\subsection{Model system}
In our model we treat the QD as a two-level system, which is justified in the
case of a strongly confined QD excited by circularly polarized light, when
only excitons with a single spin orientation can be generated. The ground
state is denoted by $|g\rangle$ and the exciton state by $|x\rangle$. The
states are split by the exciton energy $\hbar\Omega$. Taking the energy of
the ground state as zero, the system Hamiltonian $H_{\rm sys}$ reads
\begin{equation} \label{eq:h_sys}
	H_{\rm sys}=  \hbar \Omega | x \rangle \langle x |\ .
\end{equation}
The coupling to the classical light field $E(t)$ is treated in dipole and
rotating wave approximation via the Hamiltonian
\begin{equation}
	H_{\rm sys-light}= -M_0  \big[E^{(-)}(t)| g \rangle \langle x |
        + E^{(+)}(t) | x \rangle \langle g | \big]\ ,
\end{equation}
where $M_0$ is the dipole matrix element and
\begin{equation}
E^{(\pm)}(t)=\frac{\hbar\Omega_{\rm R}(t)}{2M_0} e^{\mp i \omega_{\rm L} t}
\end{equation}
are the positive (upper sign) and negative (lower sign) frequency component
of the electric field of the laser pulse with central frequency $\omega_{\rm
L}$. Here we have expressed the pulse shape in terms of the instantaneous
Rabi frequency $\Omega_{\rm R}$. Both the dipole matrix element and the Rabi
frequency are taken to be real and we assume a Gaussian envelope with
\begin{equation}
	\Omega_R(t) = \frac{\Theta}{\tau \sqrt{2\pi}}\exp\left[-\frac{(t-t_0)^2}{2\tau^2}\right]\ .
\end{equation}
The pulse duration is determined by $\tau$ and $\Theta$ denotes the pulse
area, which is defined such that in the absence of phonons a resonant $\pi$
pulse (i.e., a pulse with $\Theta = \pi$) completely excites the exciton from
the ground state.

In addition, we take into account the electron-pho\-non interaction. The phonon
Hamiltonian is given by
\begin{equation} \label{eq:h_pho}
    H_{\rm pho} =   \sum_{\mathbf{q}}   \big[\hbar\omega_{\mathbf{q}}
    b_{\mathbf{q}}^{\dag} b^{}_{\mathbf{q}} + \hbar g_{\mathbf{q}}
    \big(b_{\mathbf{q}}^{\dag}+ b^{}_{\mathbf{q}}\big) | x \rangle \langle x | \big]\ ,
\end{equation}
where $ b_{\mathbf{q}}^{\dag}$ ($b^{}_{\mathbf{q}}$) are the creation
(annihilation) operators for a phonon. For simplicity we assume that, as in
the case of bulk phonons, the phonons can be classified in terms of a wave
vector $\mathbf{q}$. A generalization to arbitrary phonon quantum numbers is
straightforward. The first part in Eq.~(\ref{eq:h_pho}) describes the free
phonon part with the dispersion relation $\omega_{\mathbf{q}}$ and the second
term describes the pure dephasing-type carrier-phonon interaction via the coupling matrix element $g_{\mathbf{q}}$. The Hamiltonian Eq.~\eqref{eq:h_pho} describes the fact that, due to the large difference between exciton and phonon energy the
phonons do not induce transitions between exciton and ground state. They
affect, however, the phase of the coherence between these states which, for
finite pulses, also has an influence on the resulting occupation. In general,
there can be different types of phonon modes (e.g., acoustic and optical,
longitudinal and transverse) and different coupling mechanisms (e.g.,
deformation potential or polar). Here we assume that all these can be treated
separately.

The operators for the phonon displacement ${\mathbf{u}}=\left<
\hat{{\mathbf{u}}}\right>$ and phonon momentum $\boldsymbol{\pi}=\left<
\hat{\boldsymbol{\pi}}\right>$ are related to the phonon mode operators
$b_{\mathbf q}^{}$ and $b_{\mathbf q}^{\dagger}$ via
\begin{subequations} \begin{equation}
	\hat{ {\mathbf{u}} }({\mathbf{r}}) = i \sum_{\mathbf{q}}
\sqrt{\frac{\hbar}{2\varrho V \omega_{\mathbf{q}}}} {\mathbf{e}}_{\mathbf{q}}
	\Bigl( b_{\mathbf{q}} + b_{\mathbf{-q}}^\dag \Bigr)
         e^{i{\mathbf{q}}\cdot{\mathbf{r}}} \label{eq:u}
\end{equation}
and
\begin{equation}
	 \hat{ {\boldsymbol{\pi}} }({\mathbf{r}}) =   \sum_{\mathbf{q}}
    \sqrt{\frac{\varrho\hbar\omega_{\mathbf{q}}}{2 V }}
	{\mathbf{e}}_{\mathbf{q}} \Big( b_{\mathbf{q}} - b_{\mathbf{-q}}^\dag \Big)
        e^{i{\mathbf{q}}\cdot{\mathbf{r}}}\ , \label{eq:pi}
\end{equation} \end{subequations}
where ${\mathbf{e}}_{\mathbf{q}}$ denotes the polarization vector of the
phonon mode, $\varrho$ is the crystal density and $V$ is the normalization
volume. In this paper, we are particularly interested in the fluctuation
properties of the phonon displacement and momentum. These are given by
\begin{equation}
    (\Delta {\mathbf{u}})^2  =\left< \hat{{\mathbf{u}}}^2 \right>- \left< \hat{{\mathbf{u}}}\right>^2
    \quad \mbox{and}
    \quad (\Delta \boldsymbol{\pi})^2  =\left< \hat{\boldsymbol{\pi}}^2 \right>
    - \left< \hat{\boldsymbol{\pi}}\right>^2\, .
\end{equation}
Squeezing occurs, if the fluctuations fall below their respective vacuum
value $(\Delta {\mathbf{u}})^2_{\rm vac} $ and $(\Delta
\boldsymbol{\pi})^2_{\rm vac}$. To simplify the discussion, we introduce the
quantities
\begin{subequations}
\begin{equation}
D_u  =\frac{(\Delta {\mathbf{u}})^2 -(\Delta {\mathbf{u}})^2_{\rm vac} }{(\Delta
    {\mathbf{u}})^2_{\rm vac} }
\end{equation}
and
\begin{equation}
D_{\pi}  =\frac{(\Delta \boldsymbol{\pi})^2 -(\Delta \boldsymbol{\pi})^2_{\rm vac} }
{(\Delta \boldsymbol{\pi})^2_{\rm vac} } \ .
\end{equation}
\end{subequations}
These definitions are particularly handy to identify squeezing, because we
only need to check if these quantities, which for simplicity we will call
fluctuations in the following, become negative. Thus, the presence of
displacement or momentum squeezing is equivalent to $D_u<0$ or $D_{\pi}<0$,
respectively.

\subsection{Generating function formalism}
To calculate the dynamics of the system, we use generating functions which
are defined as the expectation values \cite{vagov2002ele,axt2005pho}
\begin{equation}
	\rho_{\nu\nu^\prime} (\{\alpha_{\mathbf{q}}\},\{\beta_{\mathbf{q}}\})=\left<
    |\nu\rangle\langle\nu^\prime| e^{\sum_{\mathbf{q}} \alpha_{\mathbf{q}}
    b_{\mathbf{q}}^{\dag} } \, e^{\sum_{\mathbf{q}} \beta_{\mathbf{q}} b_{\mathbf{q}}^{} }
    \right> \ .
\end{equation}
Here, $|\nu\rangle$ denotes the electronic state of the system, i.e., $|\nu
\rangle \in \{|g\rangle, |x\rangle \}$ and $\alpha_{\mathbf{q}}$,
$\beta_{\mathbf{q}}$ are complex numbers. From the generating functions, all
electronic and phononic variables can be calculated. The pure phonon
variables are encoded in the function
\begin{subequations}
\begin{eqnarray}
    F(\{\alpha_{\mathbf{q}}\},\{\beta_{\mathbf{q}}\}) &&= \left< e^{\sum_{\mathbf{q}} \alpha_{\mathbf{q}}
    b_{\mathbf{q}}^\dag} \, e^{\sum_{\mathbf{q}} \beta_{\mathbf{q}} b_{\mathbf{q}}} \right> \nonumber \\
	&&= \sum_\nu \rho_{\nu\nu} (\{\alpha_{\mathbf{q}}\},\{\beta_{\mathbf{q}}\})\ ,
\end{eqnarray}
while the quantities related to the occupation of the electronic levels are
given  by the function $C$ and those related to the interband coherence are
encoded in the function $Y$ with
\begin{eqnarray}
    Y (\{\alpha_{\mathbf{q}}\},\{\beta_{\mathbf{q}}\}) &=& \left< |g\rangle\langle x|
        \, e^{\sum_{\mathbf{q}} \alpha_{\mathbf{q}} b_{\mathbf{q}}^{\dag} }
        \, e^{\sum_{\mathbf{q}} \beta_{\mathbf{q}} b_{\mathbf{q}}^{} } \right> \ , \\
    C (\{\alpha_{\mathbf{q}}\},\{\beta_{\mathbf{q}}\}) &=& \left< |x\rangle\langle x|
        \, e^{\sum_{\mathbf{q}} \alpha_{\mathbf{q}} b_{\mathbf{q}}^{\dag} }
        \, e^{\sum_{\mathbf{q}} \beta_{\mathbf{q}} b_{\mathbf{q}}^{} } \right>\ .
\end{eqnarray}
\end{subequations}
For example, for $\{\alpha_{\mathbf{q}}\}=\{\beta_{\mathbf{q}}\}=0$, we
retain the occupation $f=\left< |x\rangle\langle x| \right>=C(0,0)$, while
phononic and phonon assisted variables can be obtained by derivatives of the
corresponding functions with respect to $\alpha_{\mathbf{q}}$ and
$\beta_{\mathbf{q}}$ and setting
$\{\alpha_{\mathbf{q}}\}=\{\beta_{\mathbf{q}}\}=0$ afterwards, such as
\begin{subequations}
\begin{eqnarray}
    \langle b_{\mathbf{q}} \rangle &=& \frac{\partial}{\partial \beta_{\mathbf{q}}}
        F(\{\alpha_{\mathbf{q}}\},\{\beta_{\mathbf{q}}\})\biggl|_{\{\alpha_{\mathbf{q}}\}
        =\{\beta_{\mathbf{q}}\}=0} \ , \\
\langle b_{\mathbf{q}}^{\dag} b_{\mathbf{q}}\rangle &=&
        \frac{\partial^2}{\partial \beta_{\mathbf{q}}\partial \alpha_{\mathbf{q}}}
        F(\{\alpha_{\mathbf{q}}\},\{\beta_{\mathbf{q}}\})\biggl|_{\{\alpha_{\mathbf{q}}\}
        =\{\beta_{\mathbf{q}}\}=0} \ .
\end{eqnarray}
\end{subequations}
Using Heisenberg's equations of motion, a closed set of equations of motion
for the generating functions $F$, $Y$, and $C$ can be derived
\cite{vagov2002ele}. These are partial differential equations containing
derivatives with respect to $t$, $\alpha_{\mathbf{q}}$, and
$\beta_{\mathbf{q}}$. It has been shown that for an excitation with an
arbitrary series of ultra short pulses an analytical solution of these
equations of motion can be found, which holds for any type of dispersion
relation $\omega_{\mathbf{q}}$ and coupling matrix element $g_{\mathbf{q}}$
and thus both for optical and acoustic phonons
\cite{vagov2002ele,axt2005pho}. The limit of ultra short pulses is reached
when the pulse duration is much shorter than the characteristic
phonon-induced time scale. In this case the light-induced dynamics during the
pulses and the phonon-induced dynamics between and after the pulses can be
separated. For longer pulses, when this separation of time scales is not
anymore fulfilled, no analytical solution is known and numerical techniques
have to be applied. In the following we will concentrate on the case of
interaction with optical phonons, in which a numerically tractable set of
equations of motion for the characteristic functions can be obtained. A
detailed discussion of the phonon dynamics and phonon squeezing in the case
of acoustic phonons can be found in
Refs.~\cite{wigger2013flu,wigger2014ene,wigger2015squ}. For acoustic phonons it has been found that
squeezed single or sequences of wave packets can be generated, which travel
away from the QD with the speed of sound.

\subsection{Coupling to optical phonons}
Optical phonons are typically characterized by a negligible dispersion over
the range of wave vectors which are coupled to the QD exciton. Thus, they are
well approximated by a constant phonon frequency $\omega_{\rm LO}$. Due to their vanishing group velocity, optical phonons do not travel but stay confined to the QD region where they are generated.
Typically, longitudinal optical (LO) phonons are much more strongly coupled
to the QD exciton than transverse optical (TO) phonons. Therefore in the
following we will refer to LO phonons, although the formalism is the same for
TO phonons.

In the case of a constant phonon frequency it is possible to rewrite the
phonon modes in such a form that only a small number of modes couples to the
exciton \cite{stauber2000ele}. We call the annihilation and creation
operators in the new basis $B_{\lambda}^{}$ and $B_{\lambda}^{\dag}$. For an
$N$-level system, at most $N(N+1)/2$ modes are coupled \cite{stauber2000ele},
which in the case of a two-level system evaluates to three modes. Because we
take into account only the pure dephasing mechanism only the coupling matrix
element to the excited state is non-zero. Therefore, we can further reduce
the number of coupled modes to a single one with the coupling strength
$G=\sqrt{\sum_{\mathbf{q}} |g_{\mathbf{q}}|^2}$ \cite{reiter2011gen}. We
further define the dimensionless coupling $\Gamma=G/\omega_{\rm LO}$. The
ladder operators of the coupled mode are then given by
\begin{equation}
	B_0^{} = \sum_{\mathbf{q}} \frac{g_\mathbf{q}}{G} b_{\mathbf{q}}^{} \qquad
    \mbox{and} \qquad  B_0^{\dagger} = \sum_{\mathbf{q}} \frac{g_\mathbf{q}}{G}
    b_{\mathbf{q}}^{\dagger} \ .
\end{equation}
The other modes with $\lambda \ne 0$ are taken to be orthogonal to the
coupled one.  With this, the phonon Hamiltonian reads
\begin{equation}
	H_{\rm pho} =  \hbar\omega_{\rm LO} \sum_{\lambda} B_{\lambda}^{\dag}
    B^{}_{\lambda} + \hbar G (B_0^{\dag}+ B^{}_{0}) | x \rangle \langle x |\ .
\end{equation}
To describe the coupled exciton-phonon dynamics for this system the
generating functions can be reduced to the single-mode case with only a
single pair of variables $\alpha$ and $\beta$ according to
\begin{subequations}
\begin{eqnarray}
	F(\alpha,\beta)&=&\left< e^{\alpha B^\dagger_0 }
    \, e^{\beta B^{}_0 } \right> \ , \label{eq:theorie:F}\\
	Y(\alpha,\beta)&=&\left< |g\rangle\langle x| e^{\alpha B^\dagger_0 }
    \, e^{\beta B^{}_0 } \right> \ ,\label{eq:theorie:Y} \\
	C(\alpha,\beta)&=&\left< |x\rangle\langle x| e^{\alpha B^\dagger_0 }
    \, e^{\beta B^{}_0 } \right>\ . \label{eq:theorie:C}
\end{eqnarray}
\end{subequations}
These generating functions satisfy the closed set of equations
\begin{subequations}
\begin{eqnarray}
	i \partial_t F &=& \omega_{\rm LO}\big[\beta\partial_\beta-\alpha\partial_\alpha\big]F
    +G\big[\beta-\alpha\big]C \ ,\\
	i \partial_t Y &=& \big[ \Omega + \omega_{\rm LO}\left(\beta\partial_\beta
    -\alpha\partial_\alpha\right) + G\left(\beta+\partial_\alpha+\partial_\beta\right)
    \big] Y  \nonumber \\
		&&\qquad - \frac{1}{2}\Omega_{\rm R}(t) \big[F-2C\big]e^{-i \omega_{\rm L} t} \ , \\
	i \partial_t C &=& \big[ \omega_{\rm LO}\left(\beta\partial_\beta-
    \alpha\partial_\alpha\right) + G(\beta-\alpha) \big]C \nonumber \\
	&& \qquad - \frac{1}{2}\Omega_{\rm R}(t)
    \big[Y^T e^{-i \omega_{\rm L} t}-Y e^{i \omega_{\rm L} t} \big] \ ,
\end{eqnarray}
\end{subequations}
with $Y^T(\alpha,\beta) = Y^*(\beta^*,\alpha^*)$. To further simplify these
equations, we transform the system into a rotating frame on the polaron
shifted excitation frequency $\overline{\omega}_{\rm L}=\omega_{\rm L} -
\omega_{\rm LO}\Gamma^2$ resulting in the new variables
\begin{subequations}
\begin{eqnarray}
	\overline{F}(\alpha,\beta) &=&
	F(\alpha e^{-i\omega_{\rm LO} t},\beta e^{i\omega_{\rm LO} t})\, , \\
	\overline{Y}(\alpha,\beta)&=& \exp\left[{i\overline{\omega}_{\rm L}t+\beta\Gamma
            e^{i\omega_{\rm LO} t}}\right] \times  \nonumber \\
		&& \quad Y(\alpha e^{-i\omega_{\rm LO} t}+\Gamma,\beta
            e^{i\omega_{\rm LO} t}-\Gamma)\, ,  \\
	\overline{C}(\alpha,\beta)&=& \exp\left[{\Gamma\left(\beta
            e^{i\omega_{\rm LO} t}+\alpha e^{-i\omega_{\rm LO} t}\right)}\right]
            \times \nonumber \\
		&& \quad C(\alpha e^{-i\omega_{\rm LO} t},\beta e^{i\omega_{\rm LO} t})
\end{eqnarray}
and
\begin{equation}
	\overline{\Omega}_{\rm R} = \Omega_{\rm R} e^{i(\overline{\omega}_{\rm L}-\omega_L)t}\ .
\end{equation}
\end{subequations}
The notation can be further simplified by noticing that for the calculation of the relevant expectation values $\alpha$ and $\beta$ are not independent, instead it is sufficient to calculate the functions $\overline{f}(\alpha,t) =
\overline{f}(-\alpha^* ,\alpha,t)$ where $\overline{f}$ stands for
$\overline{F}$, $\overline{Y}$ and $\overline{C}$, respectively. The
resulting equations of motion in the case of resonant excitation read
\begin{subequations}\label{eq:eom}
\begin{eqnarray}
	\partial_t \overline{F}(\alpha) &=&  -2i\omega_{\rm LO} \Gamma {\rm Re}
        \left(\alpha e^{i\omega_{\rm LO} t}\right) \times \notag \\
	 		&&\exp\left[{-2i\Gamma {\rm Im}\left(\alpha e^{i\omega_{\rm LO} t}
        \right)}\right] \overline{C}(\alpha)\, ,  \\
	\partial_t\overline{Y}(\alpha) &=& \frac{i}{2} \overline{\Omega}_{\rm R} \Big[
        e^{\alpha\Gamma e^{i\omega_{\rm LO} t}}\overline{F}
            \left(\alpha-\Gamma e^{-i\omega_{\rm LO} t}\right) \notag\\
		 &&	- 2e^{\alpha^*\Gamma e^{-i\omega_{\rm LO} t}}\overline{C}
        \left(\alpha-\Gamma e^{-i\omega_{\rm LO} t}\right) \Big]\ , \\
	\partial_t\overline{C}(\alpha) &=& \frac{i}{2}e^{-\Gamma^2} \Big[
		\overline{\Omega}_{\rm R} e^{\alpha\Gamma e^{i\omega_{\rm LO} t}}
        \overline{Y}^* \left(\Gamma e^{-i\omega_{\rm LO} t}-\alpha\right) \notag \\
	&&	-\overline{\Omega}_{\rm R}^*e^{-\alpha^*\Gamma e^{-i\omega_{\rm LO} t}}
        \overline{Y} \left(\Gamma e^{-i\omega_{\rm LO} t}+\alpha\right) \Big]\ .
\end{eqnarray}
\end{subequations}
For the excitation with ultra short pulses, described mathematically in terms
of $\delta$-functions, an analytical solution of these equations can be found
\cite{axt1999coh,vagov2002ele,axt2005pho}, while in the case of extended
pulses, the equations cannot be solved analytically anymore. Instead, a
numerical integration of the equations of motion is needed \cite{axt1999coh},
which is feasible here since the infinite number of variables
$\alpha_{\mathbf{q}}$, $\beta_{\mathbf{q}}$ in the multi-mode case has been
reduced to a single complex variable $\alpha$. However, due to the shifted arguments on the right hand side of Eq.~\eqref{eq:eom} all $\alpha$-values are coupled. For the numerical integration we have
used a standard fourth-order Runge-Kutta method. The initial conditions were
chosen such that the system is initially in the ground state at temperature
$T=0$~K resulting in $\overline{C}(t=0)= \overline{Y}(t=0)=0$ and
$\overline{F}(t=0)=1$.

Having introduced a single LO phonon mode, it is possible to describe the
state of the LO phonons in terms of a Wigner function in the phase space
spanned by variables $U$ and $\Pi$ \cite{schleich2011qua}. $U$ and $\Pi$ are
the phase space representations of the corresponding operators $\hat{U}$ and
$\hat{\Pi}$, which are directly connected to the phonon creation and
annihilation operators of the coupled mode via
\begin{equation}
\hat{U} = B_0^{} + B_0^{\dagger} \qquad \mbox{and} \qquad
\hat{\Pi} =i( B_0^{} - B_0^{\dagger}).
\end{equation}
For simplicity, although defined in a dimensionless form, we will refer to
$U$ as displacement and $\Pi$ as momentum in the following.

The Wigner function is the quantum mechanical analogue of a phase-space
distribution function. It is a real-valued function, however, in contrast to
a classical distribution function the Wigner function can become negative.
Negative values of the Wigner function therefore indicate genuine quantum
mechanical behavior. From the generating phonon function the Wigner function
is calculated via the characteristic Wigner function \cite{schleich2011qua}
\begin{eqnarray}
	C_{\rm W}(\alpha)&& = \left< e^{ \alpha B^\dagger_0 - \alpha^* B^{}_0 }\right>
		= e^{ -\frac{1}{2}\left|\alpha\right|^2 } \left< e^{\alpha B^\dagger_0 }
    \, e^{-\alpha^* B^{}_0 } \right> \nonumber \\
	&& = e^{ -\frac{1}{2}\left|\alpha\right|^2 } F(\alpha,-\alpha^*).
\end{eqnarray}
With this, the Wigner function is
\begin{eqnarray}
	W(z)\! &=&\! \frac{1}{\pi^2}\!\iint\! e^{ \alpha^\ast z - \alpha z^\ast } C_{\rm W}(\alpha)\,
        {\rm d}^2\alpha  \\
	&=&\! \frac{1}{\pi^2}\!\iint\! e^{ \alpha^\ast z - \alpha z^\ast }
        e^{ -\frac{1}{2}\left|\alpha\right|^2 } F(\alpha,-\alpha^*)\, {\rm d}^2\alpha \nonumber \\
	&=&\! \frac{1}{\pi^2}\!\iint\! e^{ \alpha^\ast z e^{-i\omega_{\rm LO} t} -
        \alpha z^\ast e^{i\omega_{\rm LO} t}  -\frac{1}{2}\left|\alpha\right|^2 }  \overline{F}(-\alpha^\ast)\, {\rm d}^2\alpha. \nonumber
\label{eq:theorie:wigner}
\end{eqnarray}
Here, $z$ is a complex number and we define Re$(z)=\frac{U}{2}$ and
Im$(z)=-\frac{\Pi}{2}$. Non-negative probability distributions $P(U)$ and
$P(\Pi)$ can be obtained from the Wigner function by integration over $\Pi$
and $U$, respectively. From these probability distributions expectation
values and fluctuations can be calculated in the standard way such as, e.g.,
for the operator $\hat{U}$:
\begin{subequations}
\begin{eqnarray}
	\langle \hat{U}\rangle &=& \int \, U \,P(U)\, {\rm d}U  = \iint  \, U \,
    W(U,\Pi)\,{\rm d}\Pi {\rm d}U \\
    \langle \hat{U}^2\rangle &=& \int \, U^2 \,P(U)\, {\rm d}U \ .
\end{eqnarray}
\end{subequations}
Likewise, scaled fluctuations are introduced according to
\begin{subequations}
\begin{equation}
    D_U = \frac{(\Delta U)^2 - (\Delta U)_{\rm vac}^2}{(\Delta U)_{\rm vac}^2}
\end{equation}
and
\begin{equation}
    D_\Pi = \frac{(\Delta \Pi)^2 - (\Delta \Pi)_{\rm vac}^2}{(\Delta \Pi)_{\rm vac}^2}\, .
\end{equation}
\end{subequations}
Correspondingly, the phonon state is squeezed, when either $D_U<0$ or
$D_{\Pi}<0$.

\begin{figure}
\centering
  \includegraphics[width=0.5\columnwidth]{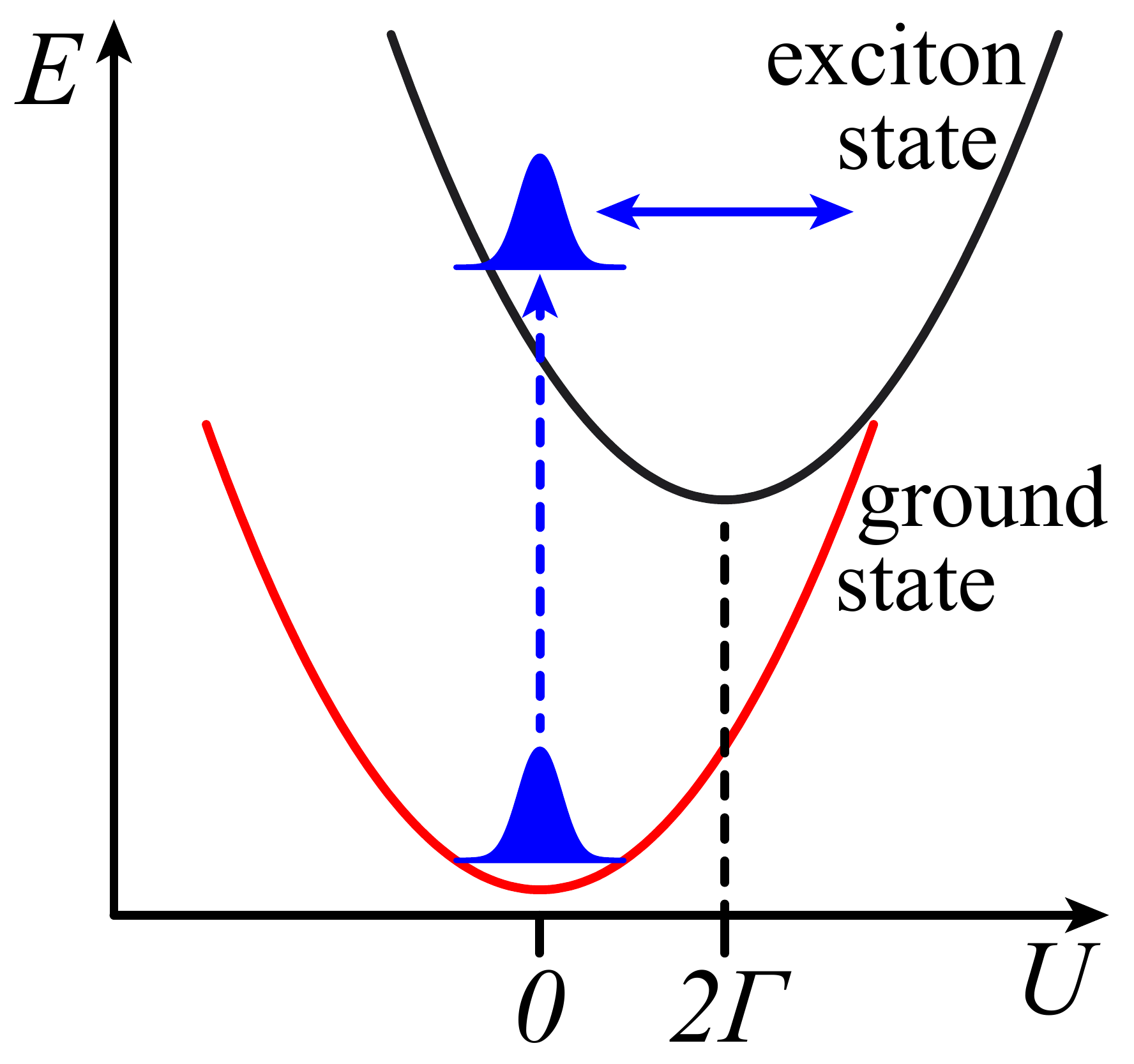}
\caption{Sketch of the phonon potential for the ground state and the exciton state.
The phonon states are marked by the blue envelopes. }
\label{fig:scheme}
\end{figure}

\section{Results} \label{sec:results}
In the following we will discuss the phonon dynamics and in particular the
possibility to achieve phonon squeezing for different excitation conditions.
We will start by discussing the limiting cases of pulses which are very short
or very long compared to the inverse of the phonon frequency. Then we will
come to the case of comparable time scales  of the light-induced and
phonon-induced dynamics.

To be specific, we consider the case of a GaAs-type QD coupled via the
Fr{\"o}hlich interaction to bulk LO phonons with an energy of
$\hbar\omega_{\rm LO} = 36.4$~meV, such that the phonon period is
$T=2\pi/\omega_{\rm LO}=114$~fs. The typical coupling strength in such dots
is rather weak with about $\Gamma=0.03$, but can be enhanced via charge
separation by applying an external electric field up to $\Gamma=0.8$
\cite{reiter2011gen}. To facilitate the interpretation of the results, in the
case of ultra short and ultra long pulses we have used an increased value of
$\Gamma=2$, which could be realized in more polar materials. A comparison
with values more typical for GaAs-type QDs can be found in
\cite{reiter2011gen}.

\subsection{Ultra short pulses}
\begin{figure}
  \includegraphics[width=\columnwidth]{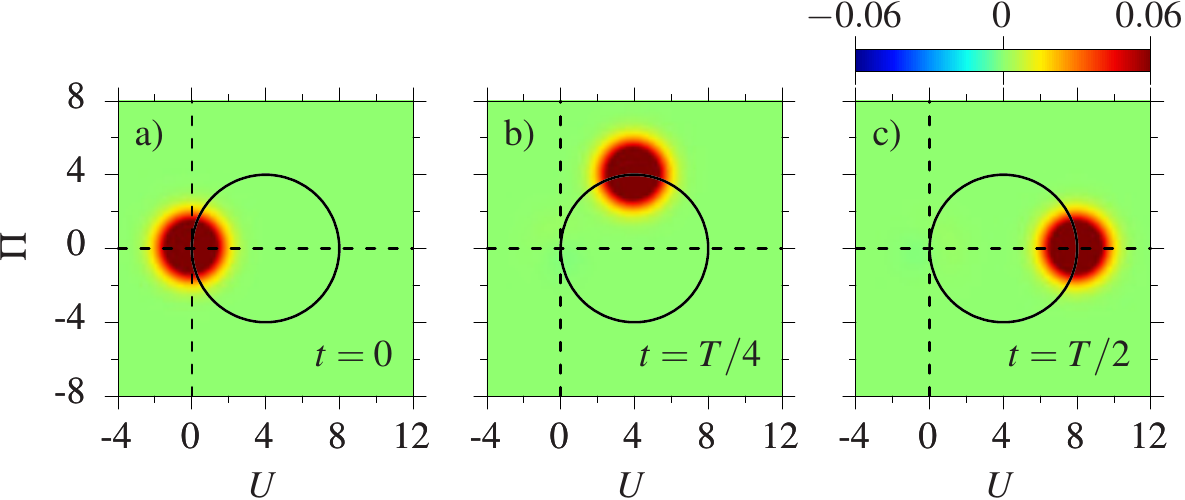}
\caption{Snapshots of the LO phonon Wigner function for a $\pi$ pulse excitation
with an ultra short pulse  ($\tau\ll T$) at three different times. }
\label{fig:limit_short}
\end{figure}

The phonon dynamics induced by a single or a pair of ultra short optical
pulses has been analyzed in detail in
Refs.~\cite{sauer2010lat,reiter2011gen}. Here we review the main results,
which serve as a reference for the case of longer pulses discussed below.

Before discussing the generation of squeezed phonons, it is instructive to
recall some features of coherent phonons. Coherent states fulfill the
Heisenberg uncertainty relation between the fluctuations of the displacement
$\Delta U$ and the fluctuations of the momentum $\Delta \Pi$ at its minimum
value. Furthermore, in the dimensionless form of $U$ and $\Pi$ introduced
above both fluctuations are equal. This condition is also realized, when the
fluctuations agree with their vacuum values. In this sense, the vacuum state
is a specific coherent state. Since phonons are described by bosonic ladder
operators in the harmonic approximation, the vacuum state can be described by
the ground state in a harmonic potential as sketched in
Fig.~\ref{fig:scheme}. Phonons are in the vacuum state, when no phonon
excitation has taken place, and in particular, when the QD is in its ground
state prior to the optical excitation. In terms of the Wigner function, the
vacuum state corresponds to a two-dimensional Gaussian with equal widths in
$U$ and $\Pi$, which is centered at the origin as displayed in
Fig.~\ref{fig:limit_short}~a). If the QD exciton is created, e.g., by the
application of a $\pi$ pulse, the harmonic potential for the phonons shifts
due to the electron-phonon interaction, thus the equilibrium position for the
phonons associated with the exciton state lies at a finite value of $U$,
which is determined by the coupling constant $\Gamma$. This is illustrated in
Fig.~\ref{fig:scheme}. In the limiting case of an ultra short pulse, the
optical excitation occurs so fast that the lattice ions cannot follow. Thus,
the potential changes instantaneously while the form and position of the wave
function are conserved. In the shifted potential, the state is now displaced
with respect to the potential minimum at $U=2\Gamma$ making it a coherent
state, which oscillates in time
\cite{janszky1992inf,vagov2002ele,sauer2010lat}. In the Wigner function for
the LO phonons shown in Fig.~\ref{fig:limit_short}~a)-c) at three different
times this is clearly visible. The Wigner function moves on a circle around
its new equilibrium position at $(2\Gamma,0)$, but keeps its form. We want to
remark that here the value of the coupling constant $\Gamma$ only determines
the position of the shifted equilibrium. The subsequent dynamics is not
affected by this value.

\begin{figure}
  \includegraphics[width=\columnwidth]{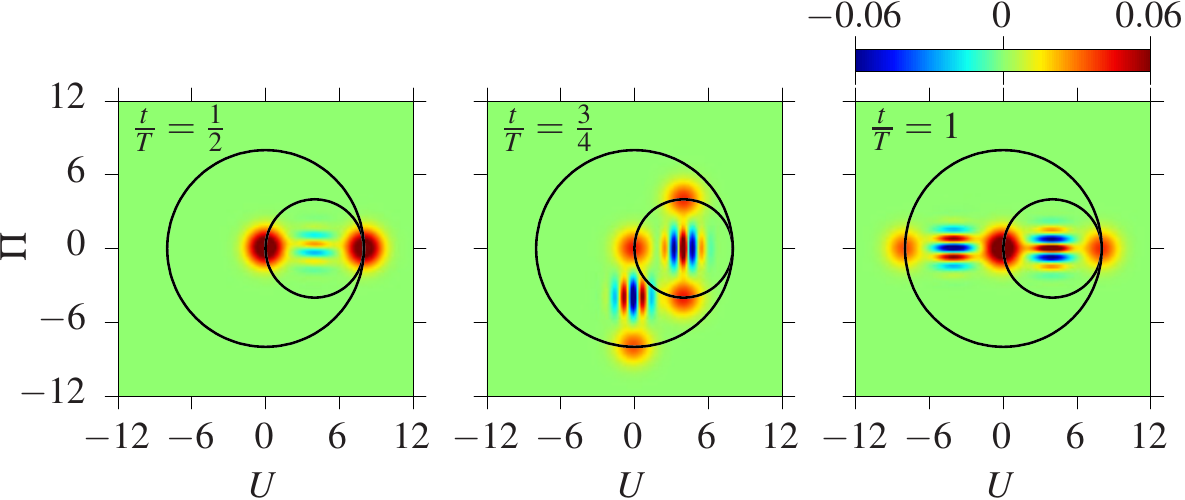}
\caption{Snapshots of the LO phonon Wigner function for an excitation with two
ultra short $\pi/2$ pulses with a delay of $T/2$ between them leading to the
formation of two cat states.}
\label{fig:cat}
\end{figure}

Let us now turn to squeezed phonons. In the Wigner function, squeezing is
seen,  when the phase space distribution looks indeed squeezed, i.e., it is
narrower in one direction in phase space than the Wigner function of the
coherent state, hence its name. There are several ways to created squeezed
states. Typically squeezing is described by the action of the a squeezing
operator on a coherent state \cite{gerry2005int}, but also special
superposition states can lead to squeezing. In the Wigner function quantum
mechanical effects are indicated by negative values, which can result in a
reduced width. In a more strict definition, squeezing occurs, if the
fluctuations of one of the variables fall below their respective vacuum
fluctuations. Because of the Heisenberg uncertainty principle this always
results in an increased fluctuation of the other variable and, thus, a
squeezed Wigner function.

The optical excitation by a single ultra short pulse only results in a
coherent state or, if the optical excitation is not complete, in a
statistical mixture of vacuum state and coherent state. Thus, squeezing is
never achieved in this case. However, using two ultra short pulses it has
been shown that it is possible to excite squeezed phonon states
\cite{sauer2010lat,reiter2011gen}. The occurrence of squeezing can be
explained by the build up of phononic cat states \cite{reiter2011gen}, which
are superpositions of two coherent states. One example of the phonon Wigner
function for the excitation with two $\pi/2$ pulses with a delay of $T/2$ is
shown in Fig.~\ref{fig:cat}. The two circles represent the movement of the
centers of the coherent states, where the smaller circle corresponds to the
exciton potential and the larger one to the ground state potential. Four
coherent states emerge, which form a mixture of two cat states seen by the
stripe pattern between two pairs of coherent states. For certain parameters
squeezing occurs in cat states due to the quantum mechanical interference
\cite{haroche2006exp}, when the states are close enough to overlap in phase
space. In Ref.~\cite{reiter2011gen} we have shown that squeezing for a two
pulse excitation can be found for a wide range of parameters considering
coupling strength, phase relation between the pulses, delay and pulse areas.

\subsection{Ultra long pulses}

\begin{figure}
  \includegraphics[width=\columnwidth]{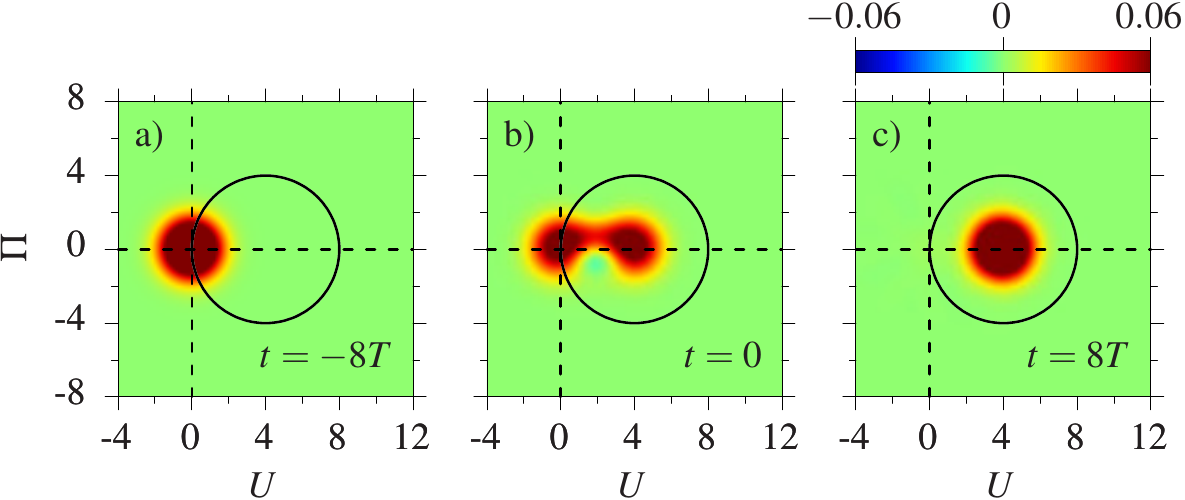}
\caption{Snapshots of the LO phonon Wigner function for a $\pi$ pulse excitation
with an ultra long pulse  ($\tau\gg T$) at three different times. }
\label{fig:limit_long}
\end{figure}

In the other limiting case of a pulse that is very long compared to the
phonon period, the situation is different. The corresponding Wigner function
is shown again at three different times in Fig.~\ref{fig:limit_long} a)-c).
In this case no analytical solution is possible, instead the Wigner function
has been obtained from the numerical solution of the equations of motion
[Eq.~(\ref{eq:eom})] for the generating functions. During the pulse the
phonons are in a mixture of the phonon ground states associated with the two
electronic potentials. Thus, the Wigner function changes its shape as seen in
Fig.~\ref{fig:limit_long} b), eventually reaching its new equilibrium
depicted in Fig.~\ref{fig:limit_long} c), as is expected for a
quasi-adiabatic state preparation of the exciton. This new equilibrium state
represents the polaron that has been built up. The final state after the
pulse is therefore again a symmetric Gaussian, but now at the new equilibrium
position $(2\Gamma, 0)$. In the exciton subsystem it is the vacuum state,
while seen from the perspective of the electronic ground state it is
displaced and thus a coherent but stationary state. Again this does not alter
the fluctuation properties of the phonons.

In Ref.~\cite{janszky1992inf} it was reported that the action of a long pulse
would result in the creation of a Fock state, which is in contrast to our
findings. We attribute this discrepancy to the use of perturbation theory in
Ref.~\cite{janszky1992inf}, which might lose its validity in the limit of
long times. Note that our generating function treatment provides  a
numerically complete solution of the dynamics in the considered model without
further approximation except for the discretization of $t$ and $\alpha$,
which is however well controlled. In particular, no perturbative
approximation is made. Indeed, for the short pulse excitation, where
perturbation theory typically works properly, our results and the results
found in Ref.~\cite{janszky1992inf} agree on the creation of a coherent
state.

From our calculations we thus conclude that both excitations with ultra short
and ultra long single pulses yield coherent phonon states after the
excitation \cite{sauer2010lat,reiter2011gen}, in the former case oscillating
around the new equilibrium position and in the latter case localized at this
equilibrium position. Also during the long pulse, the fluctuations remain at
or above the vacuum levels, as already implied by the stretched Wigner
function in Fig.~\ref{fig:limit_long} b).

\subsection{Pulses with finite duration}
We now analyze the case of excitation by pulses with pulse durations in the
intermediate regime, i.e., in the regime where the pulse duration $\tau$ and
the phonon oscillation period $T$ are of the same order. In this section we
set $\Gamma=0.5$, which is a reasonable value for a GaAs QD in the presence
of an applied electric field.

\subsubsection{Single pulse excitation}

\begin{figure}
  \includegraphics[width=0.75\columnwidth]{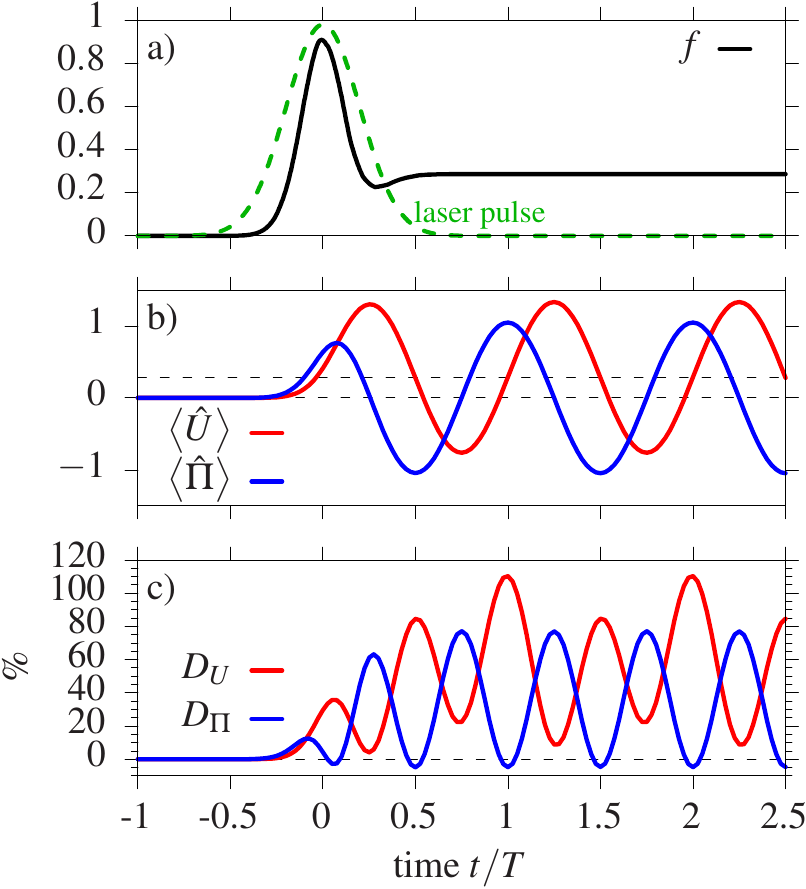}
\caption{a) Occupation of the exciton state $f$ and laser pulse, b) expectation
values of the displacement $\langle \hat{U} \rangle$ and of the momentum $\langle \hat{\Pi}
\rangle$ and c) their fluctuations. }
\label{fig:expec}
\end{figure}

Figure~\ref{fig:expec} shows the dynamics of the LO phonon system after the
excitation with a single finite pulse with a pulse area of $\Theta=2\pi$ and
a pulse length of $\tau=0.2\,T$, which in the case of GaAs corresponds to a
pulse with a full width at half maximum (FWHM) of about $50$~fs. In
Fig.~\ref{fig:expec}~a) the laser pulse and the occupation $f$ of the exciton
state is shown, while in Fig.~\ref{fig:expec}~b) the expectation values of
the displacement $ \langle \hat{U} \rangle$ and of the momentum $\langle
\hat{\Pi} \rangle$ are displayed. Figure~\ref{fig:expec}~c) shows the
respective fluctuations for these excitation conditions. During the action of
the laser pulse, which is from about $t=-0.5\,T$ to $t=0.5\,T$, the
occupation cycles once through maximum and minimum, however, due to the
phonon interaction, after the pulse an occupation of about $f=0.285$ remains
in the system. Without the interaction one would expect a final value of
$f=0$ for a pulse with $\Theta=2\pi$. Likewise the phonon expectation values
and fluctuations start to oscillate. The expectation value of the
displacement $\langle \hat{U} \rangle$ oscillates around a shifted mean
value. This shift is due to the shifted equilibrium position in the excitonic
subsystem ($2\Gamma$), which contributes to the overall mean value weighted
by the occupation $f$. So $ \langle \hat{U} \rangle$ oscillates around
$f\cdot2\Gamma=0.285$, marked as upper dashed line in
Fig.~\ref{fig:expec}~b). The fluctuations $D_U$ also show a periodic
behavior, however, for all times the fluctuations are enhanced, such that a
displacement squeezing does not take place. For the momentum, we find that
the expectation value of the momentum $\langle \hat{\Pi} \rangle$ oscillates
around $0$ and also the fluctuations $D_{\Pi}$ oscillate sinusoidal. Here, we
see each minimum is below $0$ showing the occurrence of momentum squeezing.

\begin{figure}
  \includegraphics[width=\columnwidth]{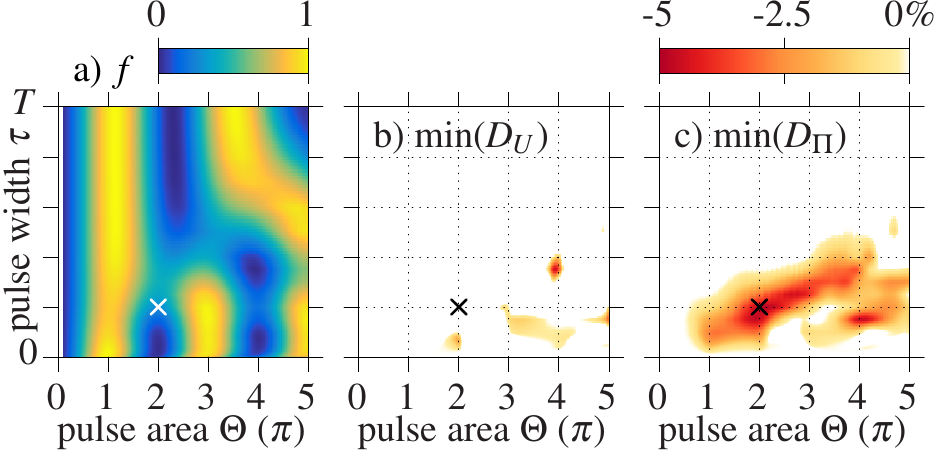}
\caption{a) Occupation of the excited state, b) minimum of the fluctuations of
the displacement and c) minimum of the fluctuations of the momentum. All at times
after the pulse and plotted as
a function of pulse area $\Theta$ and pulse width $\tau$. Note that the color
scale is restricted to negative values for b) and c). }
\label{fig:analysis}
\end{figure}

To analyze in more  detail, whether we can have squeezing for a single pulse
excitation, we have performed a systematic analysis of the fluctuations $D_U$
and $D_{\Pi}$ as function of the pulse area $\Theta$ and the pulse duration
$\tau$. For this we have extracted the minimal value of $D_U$ and $D_{\Pi}$
after the action of the pulse, e.g., for $t>0.5T$ in the previous example.
Because we consider LO phonons and we have neglected any phonon decay, the
oscillation after the pulse is periodic with the phonon period $T$ and will
go on forever. For very long times, phonon decay processes would eventually
destroy the phonon signal in real systems. Furthermore, radiative decay
processes would lead to a relaxation from the exciton to the ground state.
These processes, however, occur on much longer time scales than considered
here.

The results for the minimum values of $D_U$ and $D_\Pi$ are shown in
Fig.~\ref{fig:analysis} together with the corresponding occupation $f$ of the
exciton state. Note that for min$(D_U)$ and min$(D_{\Pi})$ we only show the
negative values for clarity, i.e., it is not shown if the minimum is above
zero. For an intermediate pulse length up to $\tau=0.5\,T$ we find that
squeezing of both displacement and momentum can indeed occur, however the
degree of squeezing is rather small with less than 5\%. We find that
squeezing occurs in the transition region between the limiting cases of ultra
short and long pulses. Interestingly, the occurrence of squeezing is
accompanied by a rather atypical behavior of the occupation. While for ultra
short pulses we find the typical Rabi rotations with maxima at
$\Theta=(2n+1)\pi$, for longer pulses, the phonon-induced renormalization of
the Rabi rotations is clearly visible, i.e., the period of the Rabi flops
gets longer. For intermediate pulse areas no clear Rabi rotations can be
identified. When we look at the fluctuations, we find that exactly in the
region with intermediate pulse duration, the fluctuations of the phonons fall
below their vacuum value, i.e., squeezed states emerge.

\begin{figure}
  \includegraphics[width=\columnwidth]{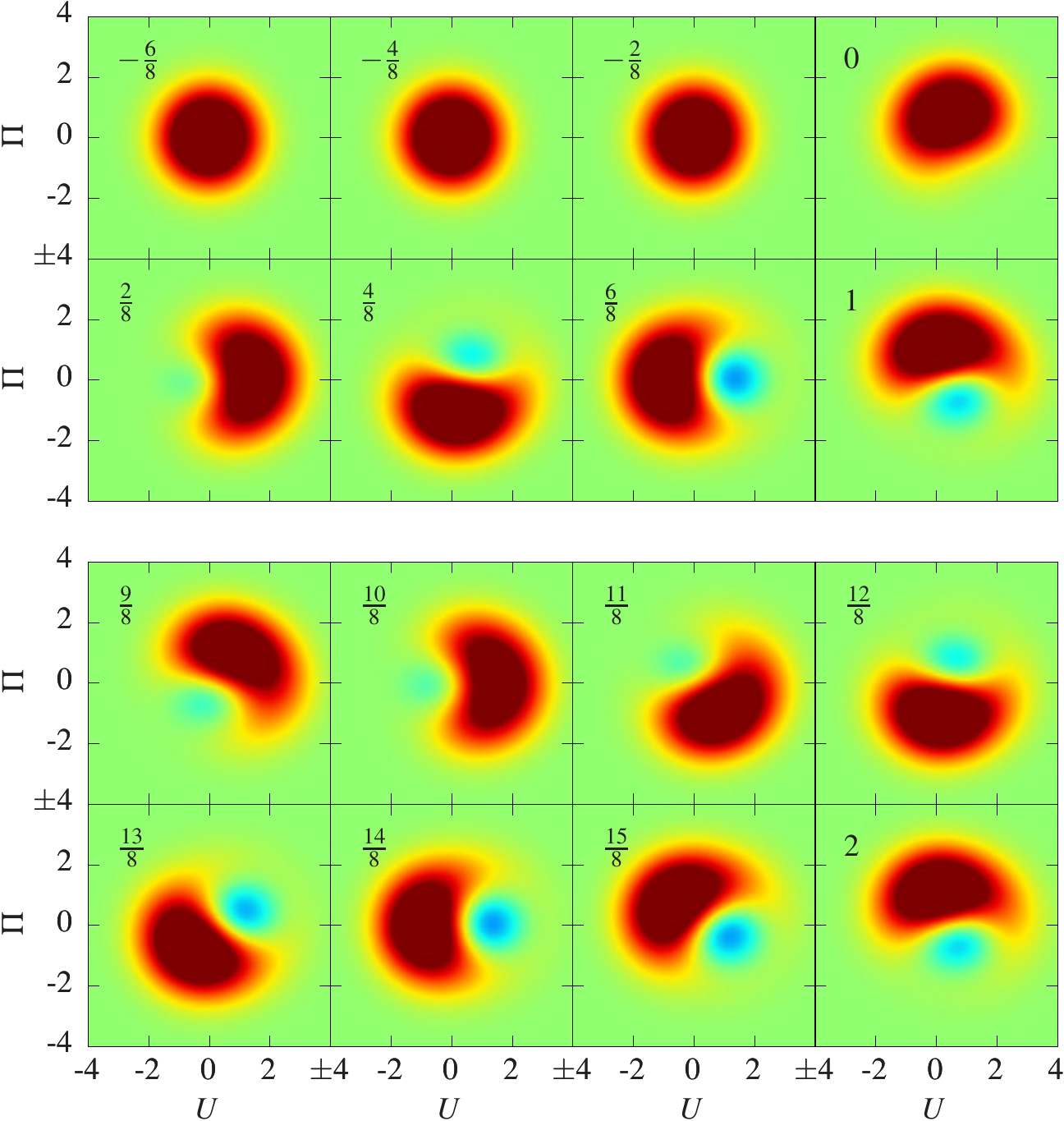}
\caption{Snapshots of the Wigner function for a $2\pi$-pulse with $\tau=0.2\,T$ for
different times $t/T$. }
\label{fig:1puls}
\end{figure}

To understand the behavior in more detail, we look at the corresponding
Wigner function plotted in Fig.~\ref{fig:1puls}. During the rise of the pulse
up to $t=0$ the phonons do not react significantly to the change in the
electronic system and essentially stay in the vacuum state. Only around
$t=0$, we find that the phonons start to noticeably react to their new
potential from the exciton state and move out of the center of phase space.
While they start to oscillate, the electronic system already moves back to
the ground state. Thus, instead of a coherent state in form of a symmetric
Gaussian, a deformed shape is formed. We also find negative values of the
Wigner function. This indicates that the phonon state contains features that
are of genuine quantum mechanical nature. After the action of the pulse, for
$t>T$, we see that the Wigner function rotates, but its form is not stable.
Around the times $t=T$ and $t=3T/2$, where the momentum is squeezed, we see
that the Wigner function is indeed elongated along $U$, but narrow in $\Pi$.
On the other hand at time $t=5T/4$ and $t=7T/4$, where the fluctuations of
the displacement have a minimum, the Wigner function is elongated along
$\Pi$. Though negative parts appear in the Wigner function, it is still wider
than in the vacuum case and no squeezing occurs.

\begin{figure}
  \includegraphics[width=\columnwidth]{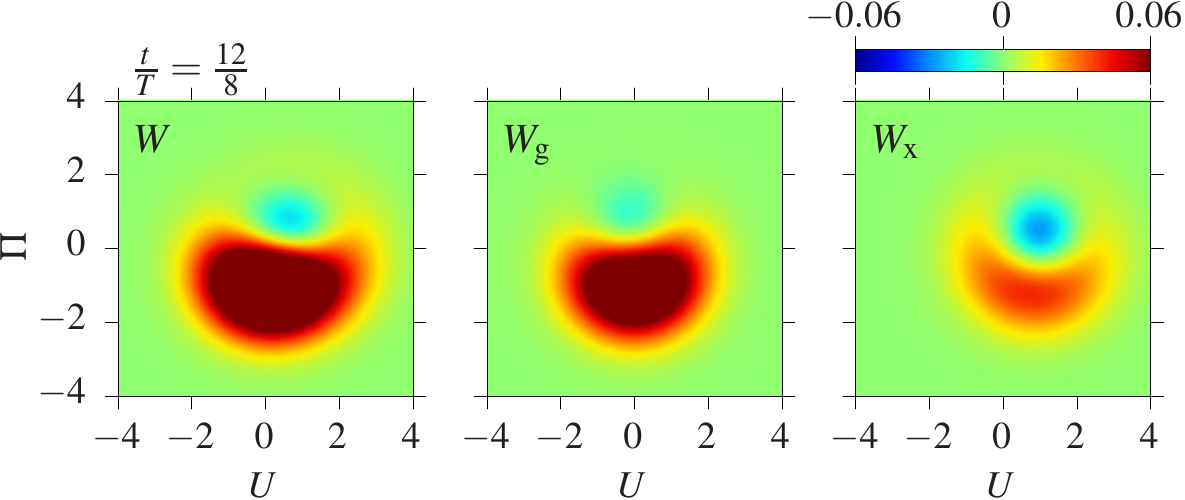}
\caption{Snapshot at time $t=3T/4$ of the Wigner function (left), which can be
separated into the Wigner function belonging to the ground state potential
$W_{\rm g}(U,\Pi)$ (middle) and the one belonging to the exciton state potential
$W_{\rm x}(U,\Pi)$ (right). }
\label{fig:GX}
\end{figure}

Since the QD is driven by a coherent light field and the initial state is a
pure state (the ground state of the QD--phonon system), the total system
consisting of QD exciton and phonons remains always in a pure state. However,
in general the temporal evolution results in an entanglement of the
electronic and the phononic subsystem. When only one of the two subsystems is
considered, the other system is traced out. This leads to a loss of
coherence. If the electronic system is traced out, the phononic system falls
apart into two parts: one belonging to the ground state potential and one
belonging to the exciton state potential. Due to the loss of coherence, these
two parts are in a statistical mixture. Note that each part on its own
corresponds to a pure state. Also the Wigner function can be separated into
these two parts with $W_{\rm g}(U,\Pi)$ being the Wigner function in the
ground state potential and $W_{\rm x}(U,\Pi)$ for the exciton potential. The
total Wigner function is the sum of both parts $W(U,\Pi)=W_{\rm
g}(U,\Pi)+W_{\rm x}(U,\Pi)$ reflecting the statistical mixture. One example
for this is shown in Fig.~\ref{fig:GX}. For both parts $W_{\rm g}(U,\Pi)$ and
$W_{\rm x}(U,\Pi)$ we find a banana shaped Wigner function consisting of a
positive region that bends around a negative one. Each part now rotates with
a stable shape around its respective equilibrium, i.e., $W_{\rm g}(U,\Pi)$
around $(0,0)$, while $W_{\rm x}(U,\Pi)$ moves around $(2\Gamma,0)=(1,0)$.
When the two parts are summed up, the shape of the total Wigner function is
not stable in time.

\begin{table}
\caption{List of probabilities $P_n^{i}$ finding Fock state $\left|n\right>$
in the subspace $i=g,x$.} \label{tab:Pn}
\begin{tabular}{lll}
\hline\noalign{\smallskip}
$n$ & $P_n^{\rm g}$ & $P_n^{\rm x}$\\
\noalign{\smallskip}\hline\noalign{\smallskip}
0 & 0.46 & 0.09 \\
1 & 0.25 & 0.19 \\
2 & 0.005 & $7\times 10^{-4}$ \\
3 & $4\times 10^{-5}$ & $5\times 10^{-6}$ \\
\hline\noalign{\smallskip}
\end{tabular}
\end{table}

The Wigner function also provides a straightforward way of expanding a given
quantum state expressed in terms of a Wigner function $W(q,p)$, be it pure or
mixed, into any basis $\left|\varphi_n\right>$. The probability of finding a
state $\left|\varphi_n\right>$ is then given by \cite{schleich2011qua}
\begin{eqnarray}
P_n &=& \Bigl< |\varphi_n\rangle \langle \varphi_n | \Bigr> \notag \\
    &=& \pi \iint W_{\varphi_n}(q,p) W(q,p)\,{\rm d}q\,{\rm d}p
\end{eqnarray}
where $W_{\varphi_n}(q,p)$ is the Wigner function representation  of the
projection operator $|\varphi_n\rangle \langle \varphi_n |$. Such an
expansion can also be done separately for the Wigner functions in the two
subspaces $W_{\rm g}(U,\Pi)$ and $W_{\rm x}(U,\Pi)$. This allows us to expand
the Wigner functions of the individual subspaces into the Fock states
$\left|n\right>$ corresponding to the respective potential. The Wigner
functions of the Fock states in the ground state subspace are given by
$$
W^{\rm g}_n(U,\Pi)=\frac{(-1)^n}{2\pi}{\rm e}^{-\frac12\left(U^2+\Pi^2\right)}
L_n\left(U^2+\Pi^2\right)
$$
with $L_n$ being the Laguerre polynomials. For the Wigner functions $W_n^{\rm
x}$  in the exciton system we take $U\to U-2\Gamma$.

As an example we expand the Wigner functions from Fig.~\ref{fig:GX}, where
pronounced  squeezing is visible. The probabilities obtained from this
expansion are listed in Table~\ref{tab:Pn}. Note that the sums of the
probabilities in each subsystem reflect the electronic occupations. In our
case, this means $\sum_n P_n^{\rm g} =
\left<\left|g\right>\left<g\right|\right>= 1-f \approx 0.715$ and $\sum_n
P_n^{\rm x}=f\approx 0.285$, as we have found in Fig.~\ref{fig:expec}~a).
Because the full state is normalized, it follows that $\sum_n P_n^{\rm
g}+\sum_n P_n^{\rm x}=1$. When looking at the actual numbers in
Tab.~\ref{tab:Pn}, we see that in each electronic subsystem only the first
two Fock states, namely $n=0$ and $n=1$, contribute significantly to the
phonon state. As the listed probabilities $P_1^{\rm g}$, $P_2^{\rm g}$,
$P_1^{\rm x}$ and $P_2^{\rm x}$ add up to almost $1$, there are essentially
no contributions from higher Fock states. The fact that the phonon state in
each subsystem forms a superposition essentially of the form
$$
c_0\left|0\right> + c_1\left|1\right>
$$
is true for a wide range of pulse areas and durations. Recently it was shown
that the photons emitted from QDs can be in a superposition of the lowest two
Fock states $|0\rangle$ and $|1\rangle$, and it was experimentally
demonstrated that these photons exhibiting squeezing \cite{schulte2015qua}.
Though, the photons couple in a different way to the QD and are described
within the Jaynes-Cummings model, this shows that squeezing in such a
superposition occurs in a wider range of systems.

\subsubsection{Two pulse excitation}

\begin{figure}
  \includegraphics[width=0.75\columnwidth]{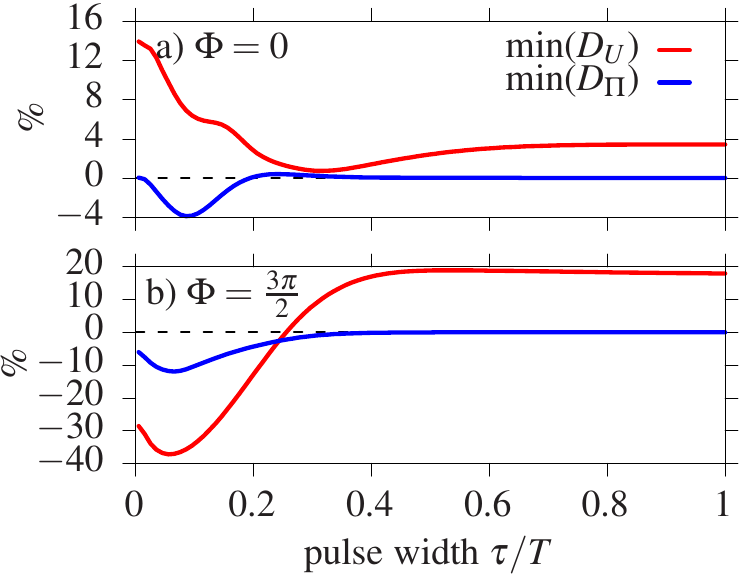}
\caption{Minimal values of the fluctuations $D_U$ and $D_{\Pi}$ for a two pulse
excitation with pulse areas $\pi/2$ each and a delay of $\Delta t=T/2$ as a
function of the pulse width. a) Phase $\Phi=0$ and b) phase $\Phi=3\pi/2$.}
\label{fig:2puls}
\end{figure}

For the two pulse excitation, we have already shown that squeezing can occur
in the case of excitation by two ultra short pulses
\cite{sauer2010lat,reiter2011gen}. The strongest squeezing emerges, when the
two pulses have a delay of $\tau=T/2$ and a pulse area of $\pi/2$ each. As we
have seen, two pairs of phononic cat states build up, which for suitable
coupling strengths (e.g., $\Gamma=0.5$) give rise to squeezing. Another
crucial pulse parameter is the phase difference $\Phi$ between the two
pulses. For a phase of $\Phi=0$ the state is not squeezed, while for
$\Phi=3\pi/2$ the squeezing is maximal. This is in agreement with squeezing
in cat states, which also depends crucially on the phase in the superposition
state \cite{gerry2005int}. Let us briefly revisit the influence of the phase
on the excitation: For the same phase, each pulse changes the occupations of
the electronic states letting them perform a part of a Rabi rotation. When
the phase difference is a multiple of $\pi/2$, the second pulse does not
change the occupation of the states, but only influences the phase difference
of the electronic states, which in turn modifies the phonon properties
drastically \cite{reiter2011gen}. In the following we will study how
squeezing prevails using finite pulses.

In Fig.~\ref{fig:2puls} we show the minimum of the fluctuations $D_U$ and
$D_{\Pi}$ for a two pulse excitation with pulse areas $\pi/2$ each and a
delay of $\Delta t=T/2$ as a function of the pulse width for two different
phases $\Phi=0$ and $\Phi=3\pi/2$. For $\Phi=0$ we do not find any squeezing
for ultra short pulses. For short pulses with $\tau=0.1\,T$ momentum
squeezing occurs with a strength of about $5$\%, similar to the case of the
one pulse excitation. For $\tau>0.2\,T$, both fluctuations are larger than
zero and we do not find squeezing. Interestingly, for $0.6\,T <\tau<T$, the
fluctuations do not change as a function of pulse width anymore. Here, the
two pulses already form a single pulse. Due to the phonon renormalization,
the pulse area is slightly smaller than $\pi$. In this case a statistical
mixture of two coherent states, one belonging to the ground state and one
belonging to the exciton state, builds up leading to increased fluctuations
in $U$, while the momentum fluctuations stay at zero. Note that these states
are stationary and do not move in time.

\begin{figure}
  \includegraphics[width=\columnwidth]{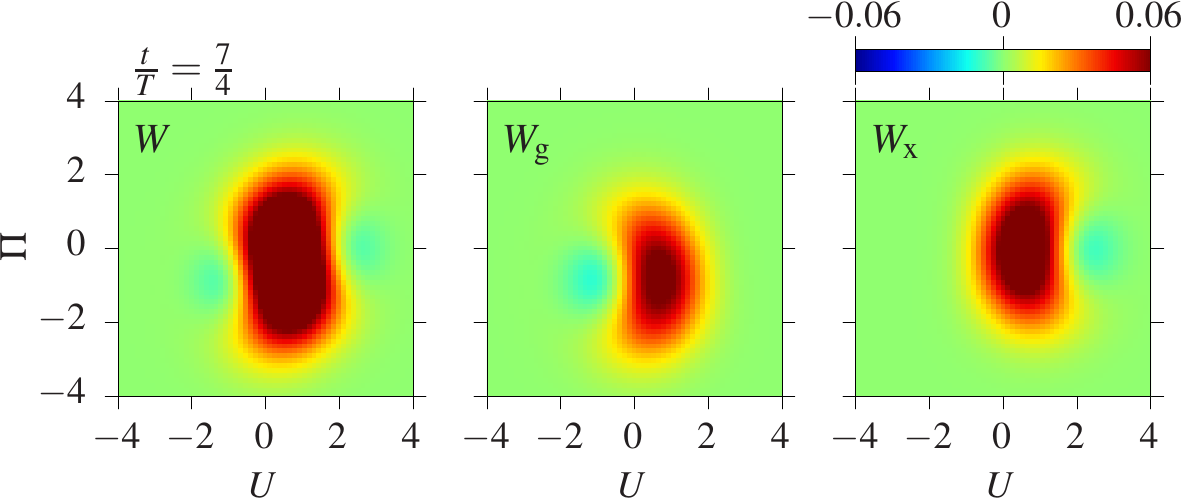}
\caption{
Snapshot at time $t=7 T/4$ of the total Wigner function $W(U,\Pi)$ (left) 
and the separation into $W_{\rm g}(U,\Pi)$ (middle) and $W_{\rm x}(U,\Pi)$ (right).
The pulse width is $\tau=0.1\,T$ for two pulses with pulse area $\pi/2$ each,
a delay of $T/2$ and a phase $\Phi=3\pi/2$.
}
\label{fig:Wig_2puls}
\end{figure}

For $\Phi=3\pi/2$ we already have squeezing of about $30$\% in the
displacement  and about $8$\% in the momentum at $\tau=0$. If now the pulse
width is taken to be finite, we see that the squeezing becomes even  more
pronounced with up to $40$\% in $D_U$ at $\tau=0.1\,T$. For larger pulse
width $\tau$ the squeezing in both displacement and momentum vanishes and for
$\tau>0.6\,T$ also reaches a constant value almost independent of the pulse
area.

It is interesting to note that also for a quantum well, i.e., for a system with a continuous electronic spectrum, a two-pulse excitation with finite pulses can lead to 
LO phonon squeezing under similar excitation conditions regarding the phase difference \cite{papenkort2012opt}. 

The occurrence of squeezing for small pulse areas can be nicely seen in the
Wigner function as displayed in Fig.~\ref{fig:Wig_2puls}, where we show a
snapshot at time $t=7T/4$, where the squeezing is maximal for the two pulse
excitation, i.e., $D_U$ is minimal. When we separate the Wigner function into
the ground state and exciton contribution with $W(U,\Pi)= W_{\rm{g}}(U,\Pi) +
W_{\rm{x}}(U,\Pi)$, we see that already in the subspaces the Wigner functions
are elongated in $\Pi$-direction with banana-like shapes oriented in opposite
directions. In addition, negative parts of the Wigner function are visible,
indicating the non-classical character of the superposition of the cat
states. When added up to the total Wigner function, as shown in the left part
of Fig.~\ref{fig:Wig_2puls}, the Wigner function becomes squeezed.

\section{Conclusions}\label{sec:conclus}
In summary, we have discussed the emergence of squeezed LO phonons by the
optical  manipulation of a QD. After reviewing the results for ultra short
excitation, we have presented new results discussing the influence of the
pulse width on the creation of squeezed phonons. To this end, we have
determined the solution of the equations of motion using a generating
function formalism, from which the Wigner function can be calculated
directly. For a single pulse, squeezing can be found if the pulse width is
about $0.05-0.5$ of the LO phonon period. In these cases superposition states
are created leading to reduced fluctuations. In the case of a two-pulse
excitation squeezing was already found for ultra-short pulses and we have
shown that a strong squeezing prevails for extended pulses up to $0.2\,T$.
Our results show that for extended pulses even stronger phonon squeezing can
be observed, which brings the theoretical predictions one step closer
to experimental realization.



\end{document}